\documentstyle[12pt]{article}

\large
\begin{document}
\setlength{\baselineskip}{21pt}
\pagestyle{empty}
\vfill
\eject
\begin{flushright}
SUNY-NTG-95-14
\end{flushright}

\vskip 2.0cm
\centerline{\bf Master Formula Approach to Chiral Symmetry Breaking:
$\pi\pi$-scattering}
\vskip 2.0 cm

\centerline{James V. Steele$^{a}$, Hidenaga Yamagishi$^{b}$, and
Ismail Zahed$^{a}$}
\vskip .4cm
\centerline{$^a$Department of Physics}
\centerline{SUNY, Stony Brook, New York 11794-3800}
\vskip .2cm
\centerline{$^b$4 Chome 11-16-502, Shimomeguro}
\centerline{Meguro, Tokyo, Japan. 153.}

\vskip 2cm

\centerline{\bf Abstract}
\vskip 3mm
\noindent
The master formula approach to chiral symmetry breaking is used to
analyze the phase shifts for $\pi\pi$ scattering in the elastic
region. The results are in excellent agreement with the data for the
phase shifts up to the $K\bar{K}$ threshold. Our analysis shows that
the $\pi\pi$ data near threshold in the scalar channel favors a large
quark condensate in the vacuum, $i.e.$ $\langle \overline q q
\rangle\sim -(240\,\,{\rm MeV})^3$.

\vfill
\noindent
\begin{flushleft}
SUNY-NTG-95-14\\
May 5, 1995
\end{flushleft}
\eject
\pagestyle{plain}
\setcounter{page}{1}

{\bf 1. Introduction}
\bigskip

Chiral symmetry offers an important framework for discussing hadronic
processes at low energy \cite{wei}.  In the past, current algebra
using the PCAC hypothesis \cite{fub} and chiral perturbation theory
using a one loop expansion \cite{gas1} have been applied to describe
processes involving pions.  Both methods are limited to the threshold
region because the former is an expansion around the soft pion point
and the latter is an expansion around the chirally symmetric
point. These points are unphysical in nature. Proposals
to remediate these shortcomings based on dispersion techniques
\cite{tru}, while well motivated from the point of view of unitarity,
do not help in understanding quantitatively the role of chiral
symmetry beyond threshold.

Recently, two of us have proposed a general framework for discussing
the role and consequences of chiral symmetry for processes involving
pions at threshold and beyond \cite{yz}.  Our approach relies on a set
of exact identities for on-shell pions, all of which following from a
master equation for the extended S matrix. This equation, which has
the structure of a reduction formula, captures the essence of chiral
symmetry and generalizes the current algebra approach on-shell.
Scattering amplitudes involving pions, nucleons and photons follow by
reduction in terms of correlation functions, some of which are
measurable. Our approach enforces chiral symmetry and unitarity to all
orders in the pion momentum, and yields exact threshold theorems that
are not necessarily contrived around the soft pion point or the chiral
point.

The purpose of this paper is to discuss in detail the $\pi\pi$
scattering amplitude as derived from the master formula in the elastic
region and compare it to the available data. The amplitude is
rewritten in terms of the scalar and vector pion form factors, as well
as vacuum correlators. The occurrence of these terms is required by
chiral symmetry and unitarity. The advantage of this approach is that
not only can these quantities be estimated in the same fashion as
chiral perturbation theory with fewer parameters, but they can also be
approximated by well accepted forms suggested by experiment without
fine tuning of the parameters or the theory. This can then be used to
assess our understanding of the low energy physics without being
hampered by the need to use a perturbation series or worry about
unitarity.

Given the importance of $\pi\pi$ scattering in hadronic physics, there
are currently a variety of theoretical approaches to $\pi\pi$
scattering most of which allow for a good deal of phenomenological
inspiration \cite{jan,au,zou}. Since our purpose is to assess the
constraints brought about by chiral symmetry alone, we will only
compare our predictions with the data and one loop chiral perturbation
theory \cite{gas1,gas2,gas3}.  For clarity, we will also display the
one loop analysis following from the master equation by power counting
in $1/f_\pi$ and minimality \cite{yz}. The latter involves only two
parameters in the isospin symmetric limit whereas chiral perturbation
theory requires four. Our results show that the master formula approach is
able to reproduce the phase shifts quite well up to the $K\bar{K}$
threshold using educated inputs for the vacuum correlation
functions.

The formulas used throughout this paper are presented in the next
section. Section 3 applies these formulas to the one loop expansion in
order to analyze $\pi\pi$ phase shifts. In section 4, the relevant
form factors are modeled in the simplest form consistent with data and
used with the master formula approach to fit phase shifts far above
the threshold region. Comparison with chiral perturbation theory
($\chi$PT) results is made throughout. Our conclusions are summarized
in section 5.

\bigskip
{\bf 2. The Calculation}
\bigskip

The master equations of the new formalism give a relation between the
scattering amplitude and various correlation functions. Taking the case
of $\pi\pi$ scattering, this relationship is quite simple.
Specifically \cite{yz}

\begin{eqnarray}
i{\cal T} \left( \pi^a\pi^b\to\pi^c\pi^d \right) &=& i{\cal T}_{\rm
tree} + i{\cal T}_{\rm vector} + i{\cal T}_{\rm scalar} + i{\cal
T}_{\rm rest} \\ &=& i\delta^{ab}\delta^{cd}A(s,t,u) + i\delta^{ac}
\delta^{bd}A(t,u,s) + i\delta^{ad}\delta^{bc}A(u,s,t) \nonumber
\label{1}
\end{eqnarray}
using the Mandelstam variables $s$, $t$, and $u$. In terms of the pion
vector form factor ${\bf F}_V$, the pion scalar form factor ${\bf
F}_S$, and the respective vector-vector and scalar-scalar correlation
functions in the vacuum, ${\bf\Pi}_V$ and ${\bf\Pi}_S$, the general
result may be written\footnote{We take $m_\pi$=140 MeV and $f_\pi$=93
MeV.} for the first three terms of equation (\ref{1}), in the form

\begin{displaymath}
A_{\rm tree}(s,t,u)={s-m_\pi^2\over f_\pi^2},
\end{displaymath}
\begin{eqnarray*}
A_{\rm vector}(s,t,u)&=&{(s-u)\over f_\pi^2}\left({\bf F}_V(t)-1-
\frac t{4f_\pi^2} {\bf\Pi}_V(t)\right) \\
&& +{(s-t)\over f_\pi^2}\left({\bf F}_V(u)-1-\frac u{4f_\pi^2}
{\bf\Pi}_V(u)\right),
\end{eqnarray*}
\begin{eqnarray*}
A_{\rm scalar}(s,t,u)&=&\frac{2m_\pi^2}{f_\pi^2}\left[-f_\pi {\bf F}_S(s) -
1 +{\langle{\hat\sigma}\rangle\over
2f_\pi}-\frac{m_\pi^2}2 {\bf\Pi}_S(s) \right]
\end{eqnarray*}
where
\begin{displaymath}
<\pi^a (p_2) | \,\,\overline{q} \gamma_{\alpha} \frac {\tau^b}2 q
(0)\,\,  |\pi^c (p_1) >
= i\epsilon^{abc} (p_1+p_2)_{\alpha} \,\,{\bf F}_V ((p_1-p_2)^2),
\end{displaymath}
\begin{displaymath}
\int d^4x e^{iq\cdot x} < 0 |
T^* \,\,\overline q \gamma_{\alpha} \frac{\tau^a}2 q (x) \,\,
\overline q \gamma_{\beta} \frac{\tau^b}2 q (0) \,\,|0>
=-i\delta^{ab} (-g_{\alpha\beta} q^2 +q_{\alpha} q_{\beta} ) \,{\bf \Pi}_V
(q^2),
\end{displaymath}
\begin{displaymath}
<\pi^a (p_2) | -\frac{\hat m}{m_{\pi}^2 f_{\pi}} \overline{q} q
(0)\,\,  |\pi^b (p_1) > =\delta^{ab} {\bf F}_S ((p_1-p_2)^2),
\end{displaymath}
and
\begin{displaymath}
\frac{{\hat m}^2}{m_{\pi}^4 f_{\pi}^2}
\int d^4x \, e^{iq\cdot x} <0| T^* \,\,\overline{q} q (x) \overline q q (0)
\,\,|0>_{\rm conn.} = -i{\bf\Pi}_S (q^2).
\end{displaymath}
The vacuum expectation value $\langle{\hat\sigma}\rangle$ was defined
in \cite{yz} to be
\begin{displaymath}
\langle{\hat \sigma}\rangle = -f_\pi-\frac{\hat m}{f_\pi m_\pi^2}
\langle {\overline q}q\rangle
\end{displaymath}
with ${\hat m}=7$ MeV being the average of the $SU(2)$ quark masses.
Taking a reasonable value for the quark condensate $\langle {\overline
q} q\rangle=-(240 \mbox{MeV})^3$ gives
$\langle{\hat\sigma}\rangle\simeq-40$ MeV. We recall that
$\langle{\hat\sigma}\rangle= 0$ corresponds to the
Gell-Mann-Oakes-Renner (GOR) relation \cite{gor}. A discussion of the
overall dependence of our result on $\langle{\hat \sigma}\rangle$ is
given at the end of section 4.

The tree amplitude was first calculated by Weinberg \cite{wei} and
must appear in any low energy theory.  Also appearing is the vector
and scalar expressions showing in a model independent way the
importance of the $\rho$ contribution \cite{weise} and the $\sigma$
state consisting of the two scattered pions \cite{ger}.  The
contribution from the ``rest'' is a correlation function of four
one-pion reduced axial vectors \cite{yz}. Since this is not a measured
quantity, its one loop form is quoted. Specifically,

\begin{eqnarray}
A_{\rm rest}^{1-loop}(s,t,u)&=&{(s-2m_\pi^2)^2\over 2f_\pi^4} \left(
{\hat c}_1 + {\cal J}(s)\right) +{(t-2m_\pi^2)^2\over 4f_\pi^4}
\left({\hat c}_1 + {\cal J}(t)\right) \nonumber \\
&& + {(u-2m_\pi^2)^2\over 4f_\pi^4}\left({\hat c}_1+{\cal J}(u)\right)
+ {\cal O}(\frac1{f_\pi^6})
\label{rest}
\end{eqnarray}
with
\begin{equation}
{\cal J}(q^2)=\frac1{8\pi^2}\left(1-\sqrt{1- {4m_\pi^2\over q^2}}
\mbox{arccoth}\sqrt{1-{4m_\pi^2\over q^2}} \right)
\label{J}
\end{equation}
representing the finite part of the pion-propagated loop.

Equation (\ref{1}) is an exact result. However, there is no known way
at present to fundamentally calculate the correlation functions or
form factors of this result.  Instead, either a loop expansion or
experimental fit must be used.  For both methods, the different
isospin contributions can be extracted from the pion scattering
amplitude in the usual manner \cite{gas1}

\begin{eqnarray*}
T^0(s,t)&=&3A(s,t,u)+A(t,u,s)+A(u,s,t)\\
T^1(s,t)&=&A(t,u,s)-A(u,s,t)\\
T^2(s,t)&=&A(t,u,s)+A(u,s,t)
\end{eqnarray*}
\begin{displaymath}
T^I(s,t)=32\pi\sqrt{s\over s-4m_\pi^2}\,\sum_{l=0}^\infty\, (2l+1)
P_l(\cos\theta) \,e^{i\delta_l^I(s)}\sin\delta_l^I(s).
\end{displaymath}
Here $\theta$ is scattering angle in the center of mass frame.  Since
elastic scattering will always be considered, no part of the amplitude
will be lost to another channel and so the phase shifts will always be
real.

To obtain the phase shift, $\delta_l^I$, an integration of the
respective $T^I$ over the Legendre polynomial with index $l$ must be
performed.  This leads to a unitarity condition on the scattering
amplitude

\begin{displaymath}
f_l^I(s)=\sqrt{1-{4m_\pi^2\over s}} \int_{-1}^1 {d(\cos\theta)\over
64\pi} T^I(s,\cos\theta) P_l(\cos\theta)
\end{displaymath}
\begin{equation}
\mbox{Im} f_l^I=\left| f_l^I \right|^2
\label{2}
\end{equation}
which serves as a consistency check on the perturbative expansion employed.

Two of the various ways to express the phase shift are

\begin{equation}
\delta_l^I(s)=\mbox{arccot}\left(\mbox{Re}\,f^I_l(s)
\over\mbox{Im}\,f^I_l(s)\right)=\frac12 \arcsin \left(2\, \mbox{Re}\,
f^I_l(s) \right)
\label{cot}
\end{equation}
The first expression is the definition for the phase shift and is
therefore preferred. However, it is sometimes better to use the second
expression as will be expounded upon below. The equivalence of these
two expressions for the phase shift may also serve as a measure of
unitarity.

\bigskip
{\bf 3. Loop expansion}
\bigskip

Calculation of the amplitude in loops is really an expansion in powers
of $s/(4\pi f_\pi)^2$.  This expansion has relevant corrections only
for $\sqrt{s}\ge 0.5$ GeV.  The amplitude is purely real at tree level
and only at one loop is there any contribution to the imaginary part.
It is therefore easy to see that such an expansion does not preserve
unitarity (eq. (\ref{2})).  For the same reason, the arcsin rather
than the arccotangent is a better approximation to the true phase
shift at one loop since two terms of the expansion are known for the
real part as opposed to only one term for the imaginary
part. Therefore the arcsin will be used for the graphs of this
section.

It is not important to quote the one loop form of the scattering
amplitude since it is worked out in \cite{yz} except to note that the
function ${\cal J}(q^2)$ given by equation (\ref{J}) appears in all
the form factors. In the $t$ and $u$ channels, the argument of $\cal
J$ will have a $\cos\theta$ dependence. The integral over the Legendre
polynomials then must be integrated numerically.

As explained in \cite{yz}, only two constants appear in the amplitude
to one loop: $c_1$ and ${\hat c}_1$. There it was shown that $c_1={\bf
\Pi}_V(0)=2f_\pi^2{\bf F}_V'(0)=0.031\pm0.001$ to one loop using the
measured value for the vector charge radius $\langle
r^2\rangle_V=6{\bf F}_V'(0)=0.42\pm0.02$ fm$^2$ \cite{exp}. In
\cite{yz}, various scattering lengths were calculated and compared
with experiment to give seven determinations of ${\hat c}_1$. The
weighted average of these values is $16\pi^2{\hat c}_1=2.46\pm0.55$.
Table 1 shows how the scattering lengths and range parameters in the
master formula approach using the middle value of ${\hat c}_1$ compare
with experiment, the tree calculation, and $\chi$PT\footnote{We used
the central values of $\overline{l}_1=-0.6$, $\overline{l}_2=6.3$,
$\overline{l}_3=2.9$, and $\overline{l}_4=4.3$ as quoted in
\cite{gas2}.} (using arcsin for all theoretical values). The phase
shifts $\delta_0^0$, $\delta_1^1$, and $\delta_0^2$ are plotted as the
solid line in figures 1, 2, and 3.  All curves in these figures cannot
be extended beyond 45 degrees because the one loop result keeps
growing like $q^2\ln q^2$ pushing the argument of the arcsin above
one.

In all three graphs, both results fit the data well up to about $450$
MeV.  The subsequent deviation is most likely an artifact of the loop
expansion, but a two loop calculation is needed to clear up the
ambiguities \cite{jim}.  The shaded region for the master equations in
$\delta_0^0$ shows the variation due to the uncertainty in ${\hat
c}_1$. This shows that up to the point of deviation, the lack of
accuracy of ${\hat c}_1$ does not matter much. Therefore the middle
value for ${\hat c}_1$ was used for all other calculations in this
paper. The master formula approach is closer to the experimental value
for $\delta_1^1$ and $\delta_0^2$ even though it has half the number
of tunable parameters compared to $\chi$PT. For $\delta_0^0$ the
master formula result is too low at threshold, but the overall shape
leads more towards the higher energy values before the $q^2$
dependence dominates. Since the one loop results in the master formula
follow from a $1/f_{\pi}$ expansion and minimality \cite{yz}, the
present results show the overall consistency of the
approach. Minimality is at the origin of KSFR \cite{ksfr}.

Taking the arccotangent form for the phase shift to one loop allows
the result to be in the full range of $0$ to $180$ degrees. It was
argued above that the imaginary part will be too small since only the
first term of the perturbation series is present to one loop.  This
would result in a lower phase shift than the full result of the theory
should give. Indeed this turns out to be the case as seen by the
dashed line in figures 4, 5, and 6. This is especially apparent for
$\delta_1^1$ which shows no sign of a $\rho$ resonance from this
analysis for either the master formula or $\chi$PT since it is
impossible to see resonances from a strict perturbative expansion due
to violation of unitarity.

\bigskip
{\bf 4. $\rho$ and $\sigma$ saturation}
\bigskip

The one loop result employed above works well near threshold but, as
was seen, cannot model even the simplest resonances. For this reason
an alternative approach would be preferred.  The master formula gives
a general result which is not dependent on the loop expansion and so
allows such an alternative. Modeling the form factors and correlation
functions that appear in eq. (\ref{1}) with reasonable forms suggested
by experiment should lead to an accurate result without much fine
tuning of the parameters or of the theory.

The only problem is that direct experimental measurements do not exist
for some of the correlation functions needed.  At least
for the vector channel it is well accepted that the features are
dominated by the $\rho$ and a good parameterization is \cite{weise,ecker}

\begin{equation}
{\bf F}_V(q^2)={m_\rho^2 + \gamma q^2 \over m_\rho^2-q^2
-im_\rho\Gamma_\rho(q^2)}
\end{equation}
with $m_\rho=770$ MeV and a momentum dependent width

\begin{equation}
\Gamma_\rho(q^2)=149\left({m_\rho\over
\sqrt{q^2}}\right)\left({q^2-4m_\pi^2 \over
m_\rho^2-4m_\pi^2}\right)^{3/2}\theta(q^2-4m_\pi^2) \mbox{ MeV}.
\label{width}
\end{equation}
For vector dominance $\gamma$ is zero. Fitting the curve to the data
for ${\bf F}_V$ requires $\gamma=0.3$ and is plotted in figure 7.
This corresponds to $\langle r^2\rangle_V=0.51$ fm$^2$ which is
reasonably close to the experimental value quoted above.

The vector-vector correlation function, ${\bf\Pi}_V(q^2)$, can be assumed
to be dominated by the $\rho$ resonance as well. Therefore the
imaginary part may be modeled by a delta function for the $\rho$ and
a theta function so that the perturbative QCD result is recovered at
large energies

\begin{displaymath}
\mbox{Im}{\bf\Pi}_V(s)=f_\rho^2\delta(s-m_\rho^2)+\frac1{4\pi}\theta(s-s_0)
\end{displaymath}
with $s_0\simeq 1.2$ GeV as a reasonable value taken for the continuum
threshold \cite{jac}.  The real part may be obtained by a once
subtracted dispersion relation.  The same can be done for the
scalar-scalar correlation function ${\bf\Pi}_S(q^2)$ with a twice
subtracted dispersion relation due to the form of the perturbative QCD
result. Smearing out the delta function by giving a width to the
resonances, yields

\begin{eqnarray*}
{\bf\Pi}_V(s)&=&\frac{f_\rho^2}{m_\rho^2-s-im_\rho\Gamma_\rho} -
\frac1{4\pi^2} \ln\left(1-{s\over s_0}\right) +
{\bf\Pi}_V^{\mbox{pert.}}(0)\\
f_\pi^2\,{\bf\Pi}_S(s)&=&\frac{f_1^{\ 2}}{m_1^{\ 2}-s-im_1\Gamma_1} -
\frac{3{\hat m}^2}{4\pi^2 m_\pi^4}\, s \,\ln\left(1-{s\over
s_0}\right)\\
&&+f_\pi^2\left({\bf\Pi}_S^{\mbox{pert.}}(0)+
s {\bf\Pi}_S^{\mbox{pert.}}\ '(0)\right).
\end{eqnarray*}
Additional resonances can be added to the scalar channel if needed by
adding further terms similar to the first term in the equation for
${\bf\Pi}_S$.  Vacuum dominance in the vector channel requires
${\bf\Pi}_V^{\mbox{pert.}}(0)=0$, however this requirement can be
relaxed for scalar particles and used to enable a better fit to
data.  Therefore let $\alpha\equiv
f_\pi^2{\bf\Pi}_S^{\mbox{pert.}}(0)$ and $\beta\equiv
f_\pi^4{\bf\Pi}_S^{\mbox{pert.}}\ '(0)$ which are both dimensionless.  The
widths are zero for spacelike momenta and are taken to be
eq. (\ref{width}) for the vector width and

\begin{displaymath}
\Gamma_1(q^2)=\Gamma_1\left({1-4m_\pi^2/q^2\over
1-4m_\pi^2/m_1^{\ 2}}\right)^{1/2}
\end{displaymath}
for the scalar.

Expanding the resonance forms in $s$ and matching coefficients with
those of the one loop result \cite{yz} in the vector channel gives
$f_\rho=(140\pm4)$ MeV in agreement with the experimental value of
$144$ MeV. Furthermore, if one resonance is assumed in the scalar
channel, matching threshold parameters for the scalar channel gives
$\alpha=-2\times10^{-2}$, $\beta=-5\times10^{-4}$, and $f_1=125$
MeV. These values were used for the fits to the phase shifts and are
close to those used in \cite{jac} (which in our notation would be:
$m_1=540$ MeV, $f_1=90$ MeV, and $s_0=1.1$ GeV).

Finally considering the scalar form factor, ${\bf F}_S(q^2)$, there is
not much to hint at its form. At tree level \cite{yz} ${\bf
F}_S(0)=-1/f_\pi$, so a reasonable form assuming the same one
resonance as used for ${\bf\Pi}_S$ is

\begin{displaymath}
{\bf F}_S(s)= -\frac1{f_\pi}{h_1m_1^2\over m_1^2-s-i m_1\Gamma_1}.
\end{displaymath}
Using the empirical determination for the scalar charge radius
\cite{gas1} $6 {\bf F}_S'(0)/{\bf F}_S(0)=0.7\pm0.2$ fm$^2$ \cite{yz}
and taking $h_1=1$ gives $m_1= 580$ MeV. Fitting to the data for
$\delta_0^0$ requires the only free parameter, $\Gamma_1$, to be $175$
MeV. Figure 7 shows that $|{\bf F}_V|$ and ${\bf\Pi}_V$ both compare
well with the data \cite{exp}. The imaginary part of ${\bf F}_S$
normalized by ${\bf F}_S(0)$ is comparable to a dispersive analysis
\cite{au} shown by the dashed line.

Even though there is no definite resonance below the $f_0(980)$, many
have interpreted the large enhancement of the $\delta_0^0$ data in the
region of $\sqrt{s}=500-800$ MeV (see figure 4) to be due to a
$\sigma$ ``particle'' with mass around $500$ MeV and width $\approx
700$ MeV \cite{ger}. This scenario is consistent with our value for
$m_1$ but our $\Gamma_1$ is much smaller. It should be noted that in
the master formula approach a large contribution from the $t$-channel
$\rho$ resonance is also seen. This is in agreement with conjectures
set forth by other authors \cite{zou} although it is not sufficient to
alone fit the data in this region.  Figure 4 shows that
$(m_1,\Gamma_1) = (580,175)$ MeV reproduces the data quite well up to
about $650$ MeV. Above this the fit rises slower than the data.

Enhancement of the $\delta_0^0$ resonance at higher energies and also
a leveling of the sharp peak in ${\bf\Pi}_S$ (figure 7) can be done
through the addition of further resonances. It was found that
including both of the next two $0^+(0^{++})$ resonances: $f_0(980)$
and $f_0(1300)$ worked well. The fit to the data is shown by the solid
line in figure 8 and the form factors in figure 9.  Values for the
constants employed are shown in table 2. The result strongly depends
on the values of the $h_i$'s and only weakly depends on the $f_i$'s.
Since the first resonance at $560$ MeV fit the data very well, its
constants were retained.  For such a simple model of the resonances
the fit is quite good up to the $K\bar{K}$ threshold. Here the strange
particles' contribution and the role of the ``rest'' term, two
problems which may be related, need to be taken into account and will
be studied elsewhere \cite{su3}. This fit shows in ${\bf\Pi}_S$ (see
figure 9) that even a narrow resonance around $580$ MeV can be
obscured by wide neighboring resonances. Further analysis is required.

The results are very good in the vector channel as well, giving the
$\rho$ resonance as expected in figure 5. This is primarily due to
three reasons: there is a large amount of data available in this
channel and so the model used for the form factors is
phenomenologically based, there is only one resonance of importance in
the energy region of interest, and the ``rest'' does not contribute
much to one loop showing this is probably the case in general. In
addition the scalar part of the amplitude has only a small
contribution and hence the one resonance or the three resonance scalar
fit discussed above give identical results. For definiteness, one
scalar resonance was used.

The ``rest'' term is important to the $\delta_0^2$ result. In fact it
is the only term which significantly contributes to the imaginary part
of the scattering amplitude. This term was not taken into account in
the resonance forms and so the phase shift is nearly always $0^\circ$
as seen by the solid line in figure 6.  However, using the one loop
expression for the ``rest'' is justified up to about $0.5$ GeV as seen
in the last section. Using the one resonance fit and including the one
loop ``rest'' agrees very well with data and is also plotted in figure
6 as the dashed line. The imaginary part of the expanded ``rest''
dominates at high momenta due to the $q^2$ dependence and hence the
result begins to deviate around $450$ MeV.  Also figure 6 includes the
arccotangent form of the $\chi$PT result given by the dotted-dashed
line.

To see the threshold behavior, the combination $\delta_0^0-\delta_1^1$
is plotted for the master formula approach using one resonance and for
both the master formula approach and $\chi$PT in the one loop
approximation in figure 10. All three give comparable results near
threshold where there is data. This combination is no longer as
important a parameter to plot since detailed information on both phase
shifts independently is now available.

As discussed above, the value of $\langle{\hat\sigma}\rangle=-40$ MeV
was used in all calculations. However, the GOR relation gives
$\langle{\hat\sigma}\rangle=0$.  Furthermore, if the quark condensate
$\langle {\overline q}q\rangle$ is zero then
$\langle{\hat\sigma}\rangle = -93$ MeV.  The dependence of the master
formula solutions on the value of $\langle{\hat\sigma}\rangle$ can be
seen best in the scalar channel which contains the largest
contribution. Only the real part of the scattering amplitude is
affected. Figure 11 shows the dependence of $\delta_0^0$ on these
three values of $\langle{\hat\sigma}\rangle$ for the one loop master
formula calculation. This graph shows that
$\langle{\hat\sigma}\rangle=-93$ MeV, corresponding to a zero quark
condensate, has the wrong shape at threshold. The other two values for
$\langle{\hat\sigma}\rangle$ are comparable with the data. Figure 12
shows how the combination $\delta_0^0-\delta_1^1$ changes with the
three different values of $\langle{\hat\sigma}\rangle$. Here the
$-40$~MeV value seems to fit best.

As was pointed out by \cite{CP}, the difference
$\delta_0^0-\delta_0^2$ at the $K^0$ mass ($498$ MeV) is a direct
measurement of CP violation. The most recent data \cite{data2} gives
$29.2\pm3^\circ$ from $\pi p\to \pi\pi\Delta$. A more direct
measurement from $K\to\pi\pi$ decay is not available, although
analyses based on chiral perturbation theory have been carried out
\cite{don}.  A weighted average of the data prior to $1975$ gives
\cite{dickey} $41.4^\circ\pm8.1^\circ$. Using the master formula
resonance result (with one scalar resonance) gives $29.8^\circ$. Use
of the one resonance result with the one loop ``rest'' added is not
reliable at this energy but gives $53.0^\circ$. The one loop expansion
in general can only be used in the arccotangent form since the arcsin
form of $\delta_0^0$ does not extend to the $K^0$ mass. This gives
$37.2^\circ$. In comparison, the arccotangent form for $\chi$PT gives
$48.3^\circ$ (to be specific, \cite{gas2} quote $45^\circ\pm6^\circ$
where they use the arccotangent with the real part only).

\bigskip
{\bf 5. Conclusions}
\bigskip

The master formula approach to $\pi\pi$ scattering offers powerful
constraints on the scattering amplitude solely on the basis of chiral
symmetry and unitarity. As a loop expansion, it fits the threshold
region as well as chiral perturbation theory with half as many
parameters to fix, and is compatible with KSFR. As a resonance fitting
of the form factors and the vacuum correlation functions, it is in
excellent agreement with the data up to the $K\bar{K}$ threshold. We
stress that the various form factors and correlation functions are
amenable to fundamental estimates in QCD or empirical
measurements. With a minimum number of assumptions, this approach
provides a testing ground for an empirical understanding of scalar and
vector correlators from the $\pi\pi$ data. Also, the $\pi\pi$ data
near threshold in the scalar channel, although not very accurate,
suggests that the vacuum supports a large quark condensate, and points
in favor of the GOR relation. Better measurements at threshold will
settle these important theoretical issues.

The present results also show that chiral symmetry constraints go far
beyond the threshold region. In particular, we have found that in the
vector channel the $\rho$ is dominant, while in the scalar channel the
presence of a low-lying though not necessarily broad resonance is
unavoidable. Since our approach shows clearly the interplay between
correlations in the vacuum and measured scattering amplitudes, it is
well suited for lattice estimations.  Improvements can even be made to
the master formula approach by an expansion to $SU(3)$ and inclusion
of kinematical nucleons. These additions should allow a much better
understanding of the inelastic region and will be carried out
elsewhere \cite{su3}.

\bigskip
{\bf Acknowledgements}
\bigskip

This work was supported in part by the US DOE grant DE-FG-88ER40388.

\newpage

{\bf Figure Captions}
\bigskip

\begin{itemize}
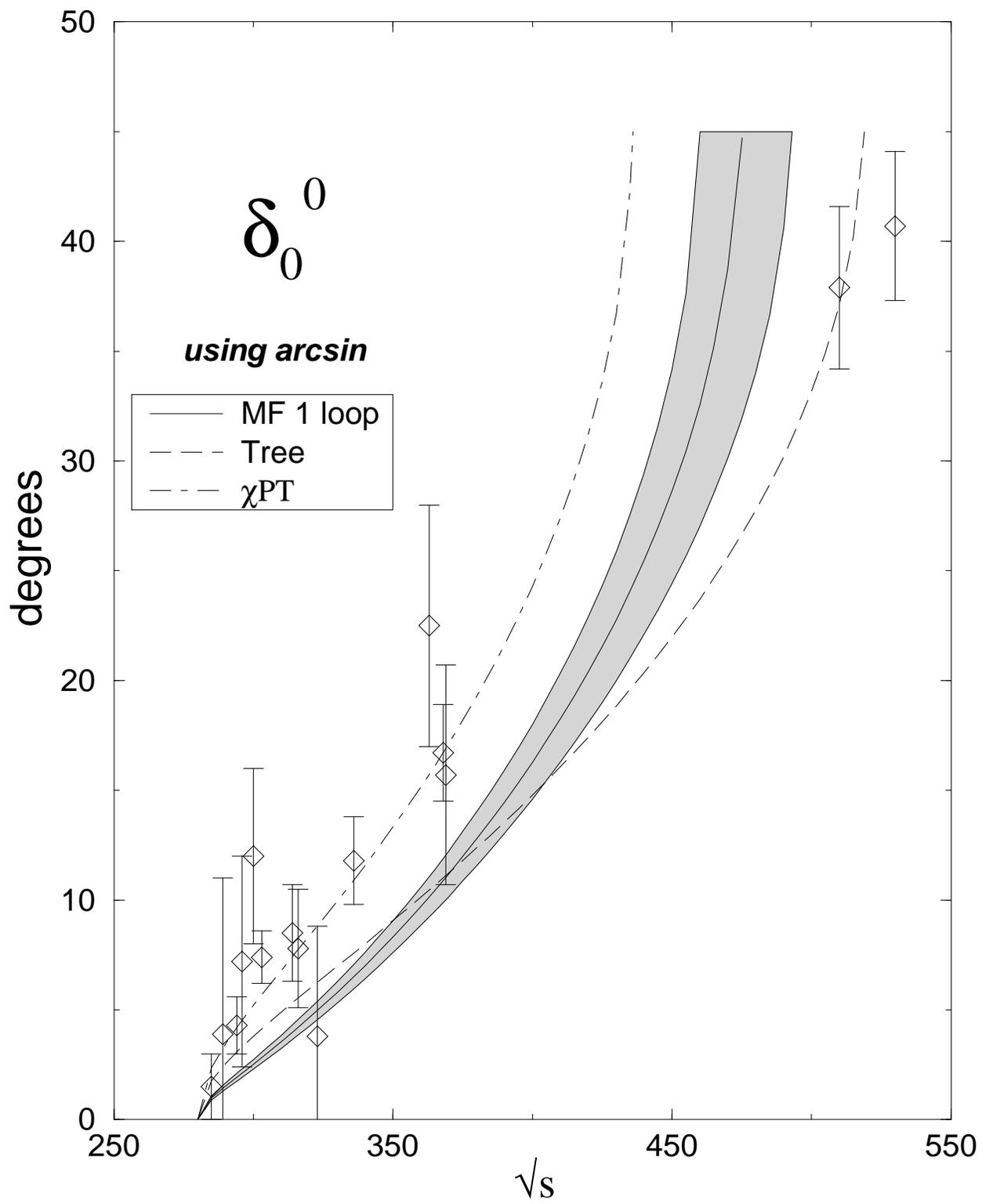
\item Fig.1. The $\delta_0^0$ phase shift using the arcsin for the one
loop expansion. The shaded region is the master formula result taking
into account the uncertainty in ${\hat c}_1$. The central value of
$16\pi^2{\hat c}_1=2.46$ is given by the middle solid line. The tree
result is shown by the dashed line and the dotted-dashed line is the
$\chi$PT result. Data is from \cite{data} below $400$ MeV and
\cite{esta} otherwise.
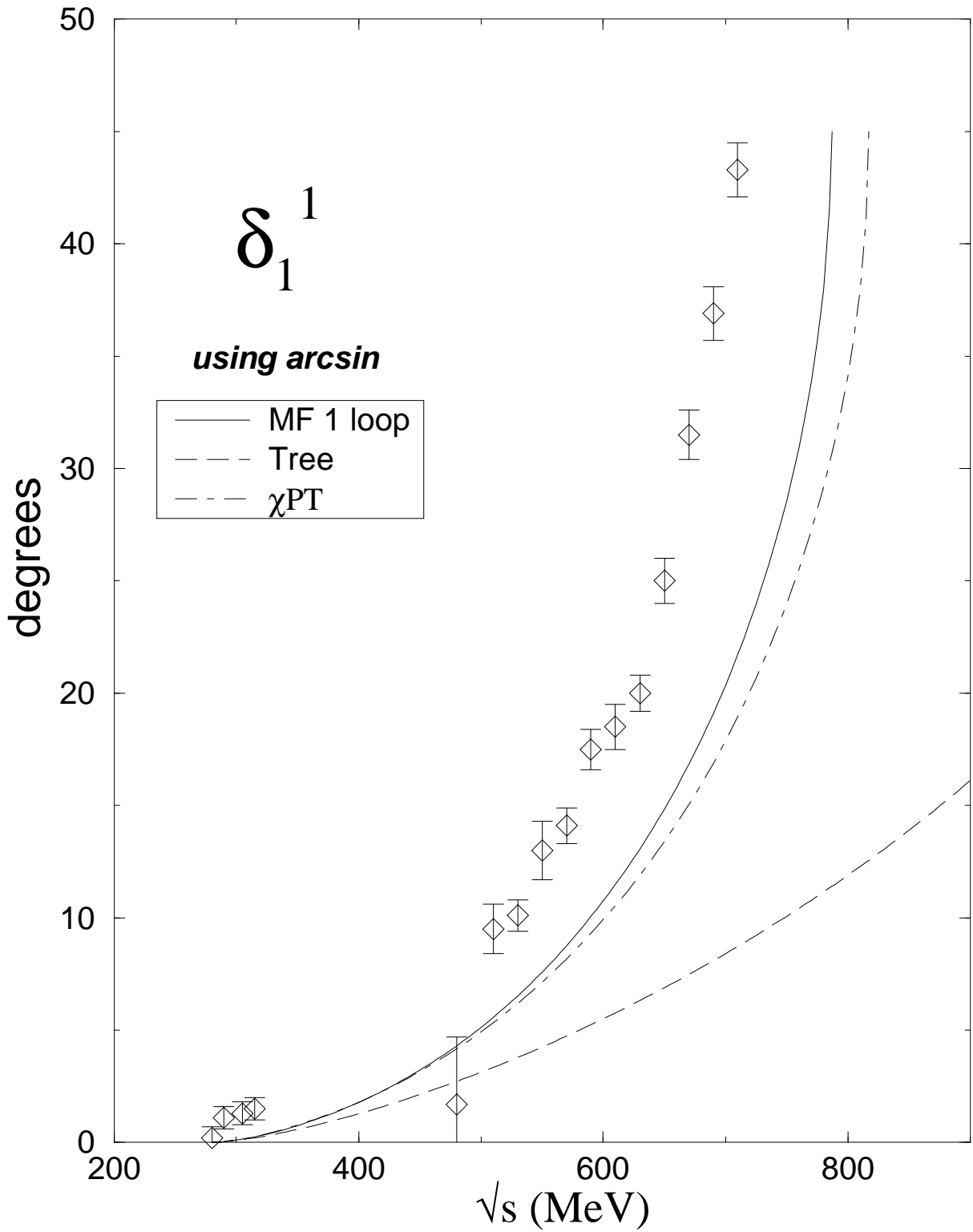
\item Fig.2. The $\delta_1^1$ phase shift using the arcsin for the one
loop expansion. The solid line is the master formula results, dotted
line is the tree result, and the dotted-dashed line is the $\chi$PT
result. Data is from \cite{data,esta}.
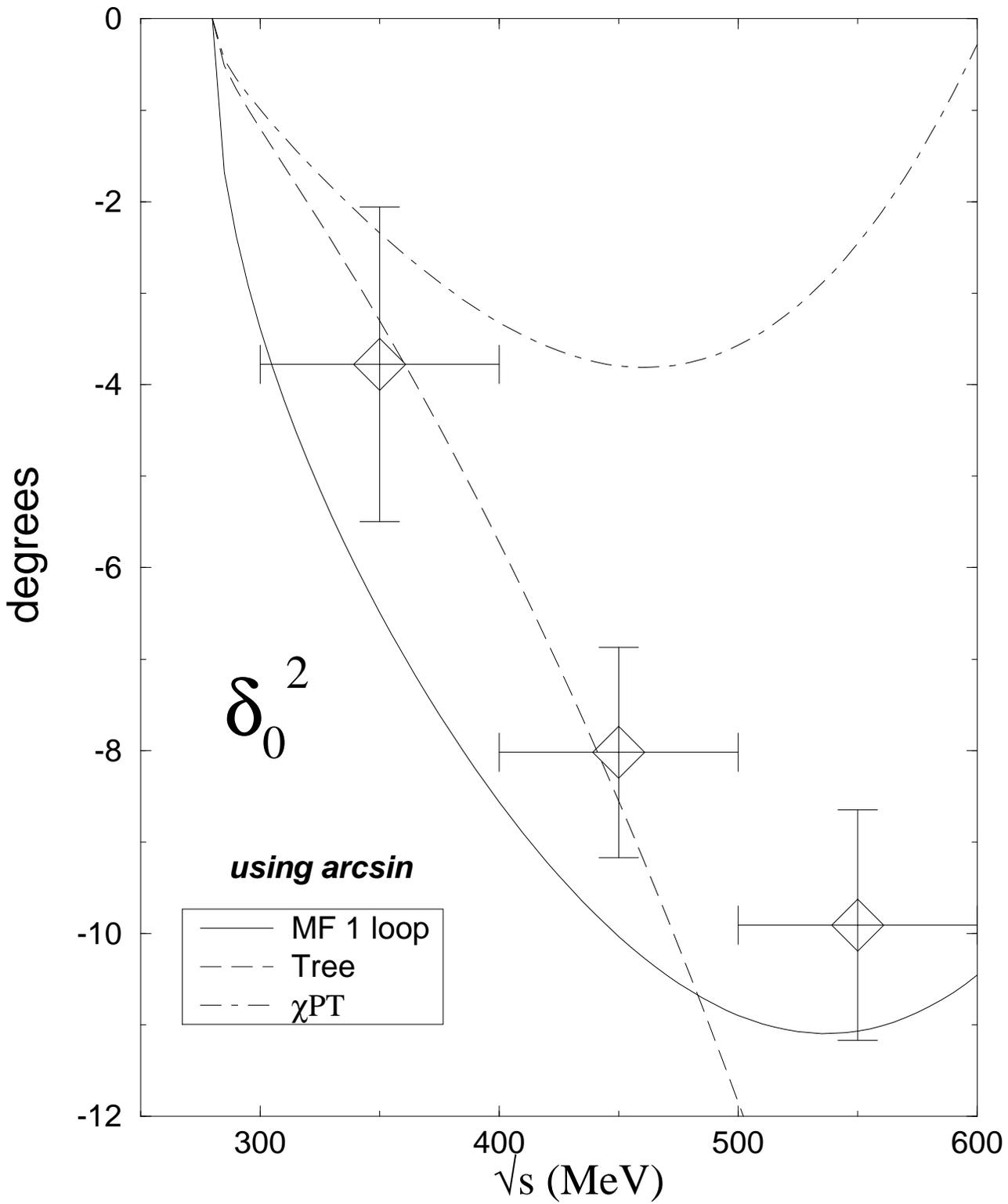
\item Fig.3. The $\delta_0^2$ phase shift using the arcsin for the
master formula (solid line), tree (dashed line), and $\chi$PT
(dotted-dashed line). Data is from \cite{hoog}.
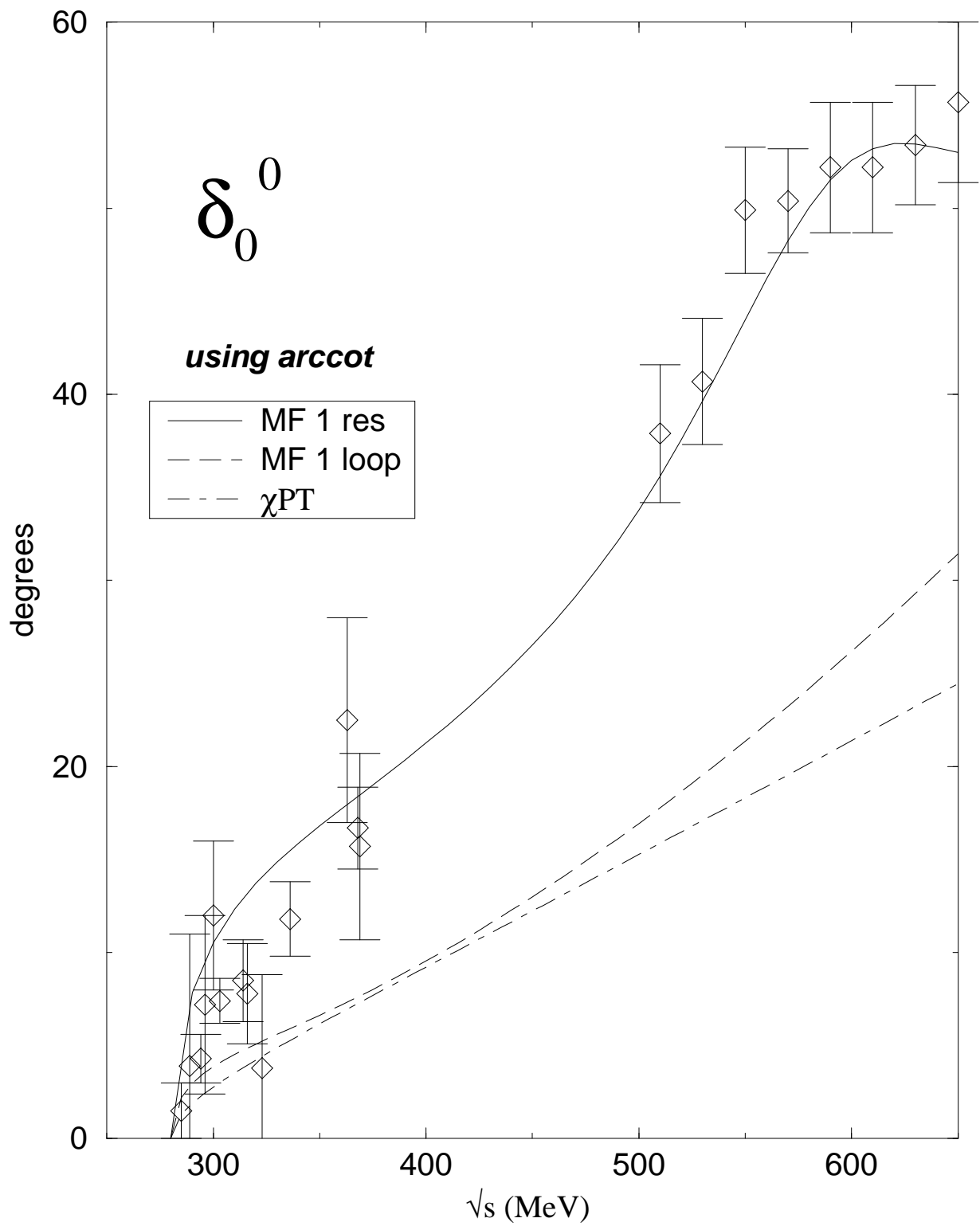
\item Fig.4. $\delta_0^0$ using the arccotangent. Here the solid line is
the one resonance fit of the master equations for one resonance, the
dotted-dashed line is $\chi$PT to one loop, and the dashed line is the
master formula result to one loop. Data is from \cite{data,esta}.
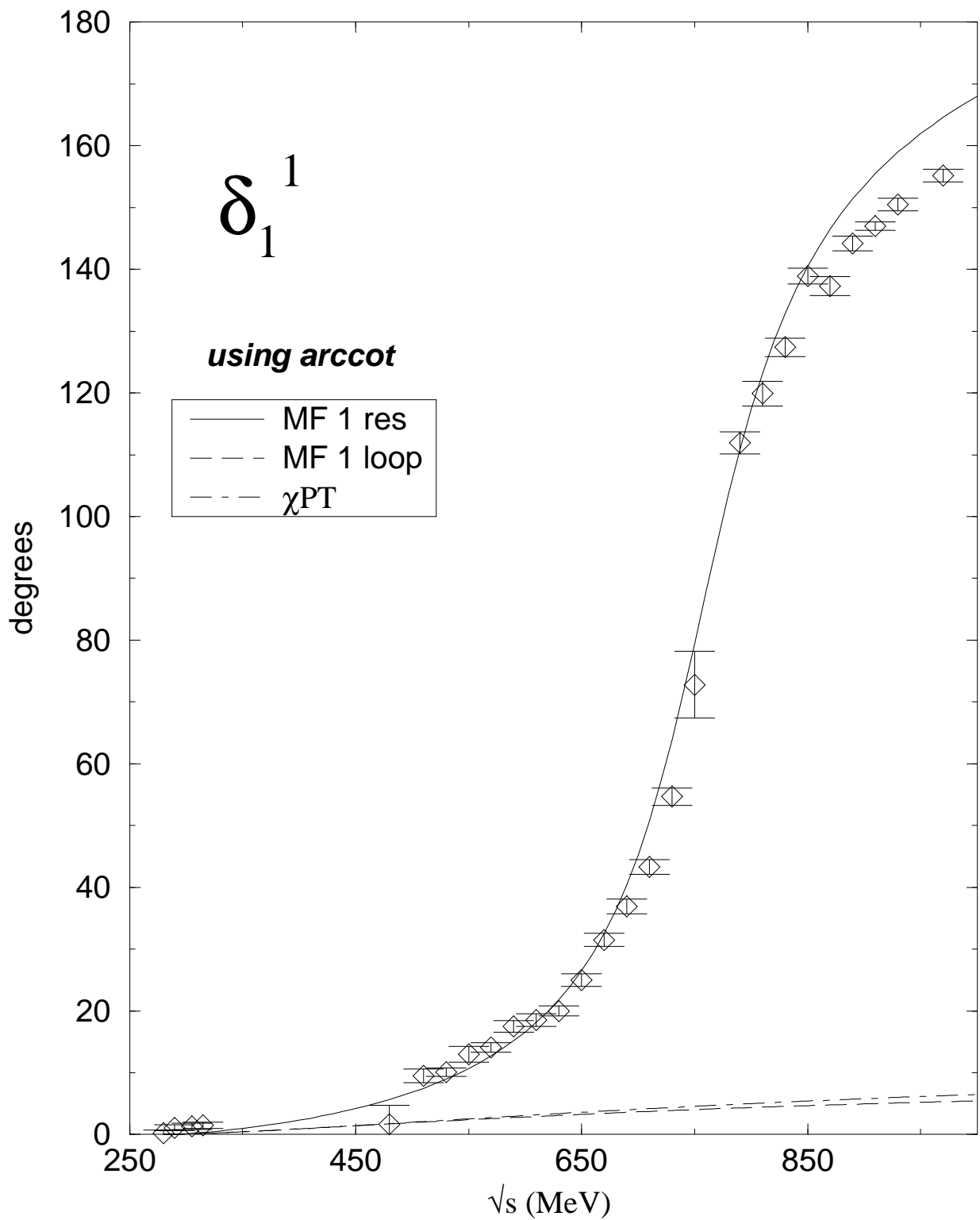
\item Fig.5. $\delta_1^1$ using the arccotangent for the master
formula with one resonance (solid line), master formula expanded to
one loop (dashed line), and $\chi$PT (dotted-dashed line). Data is
from \cite{data,esta}.
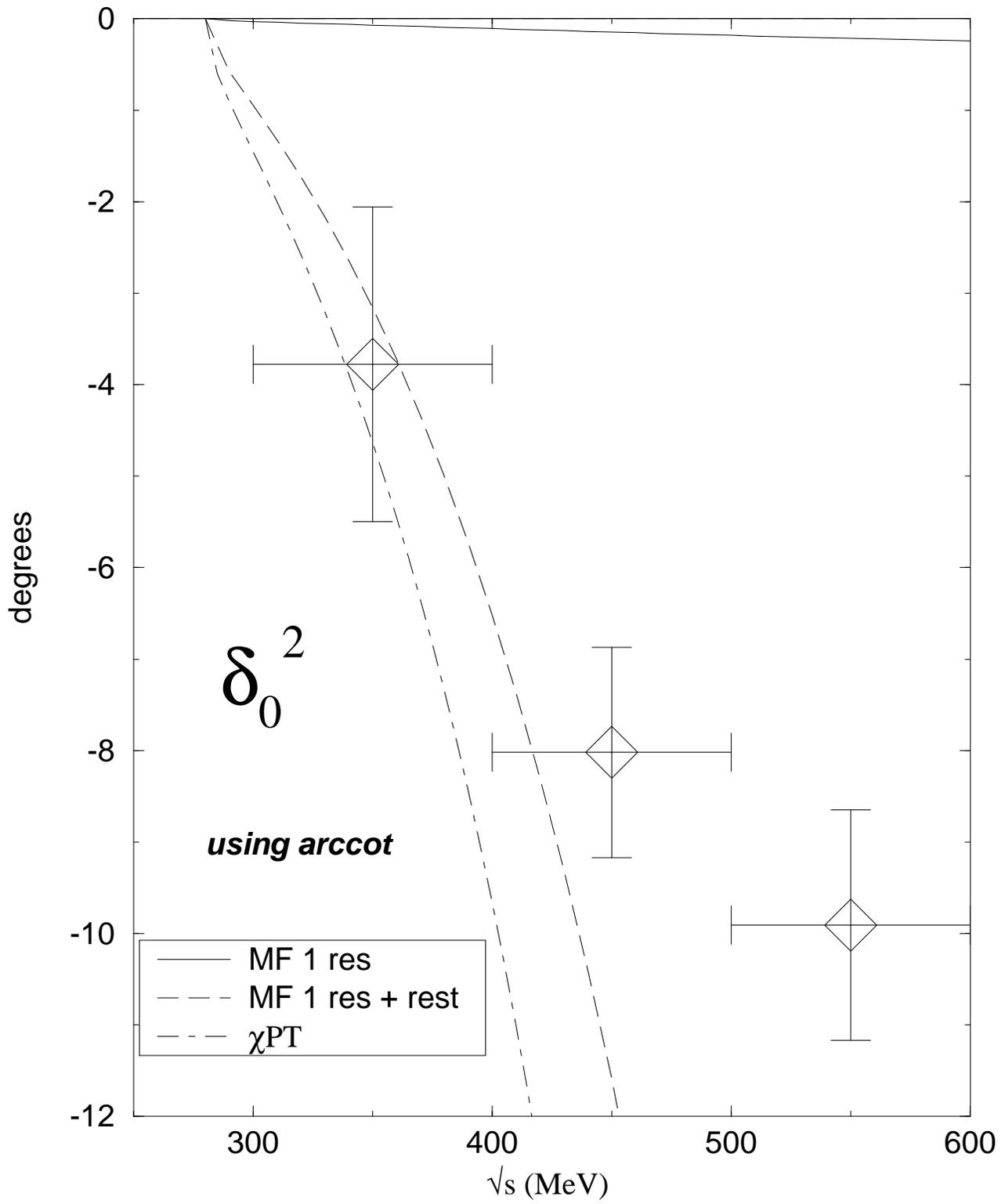
\item Fig.6. $\delta_0^2$ using the arccotangent for the master formula
approach using one resonance (solid line) and the one loop expanded
``rest'' along with the one resonance (dotted line). The $\chi$PT
result using arccotangent (dashed line) is also shown. Data is from
\cite{hoog}.
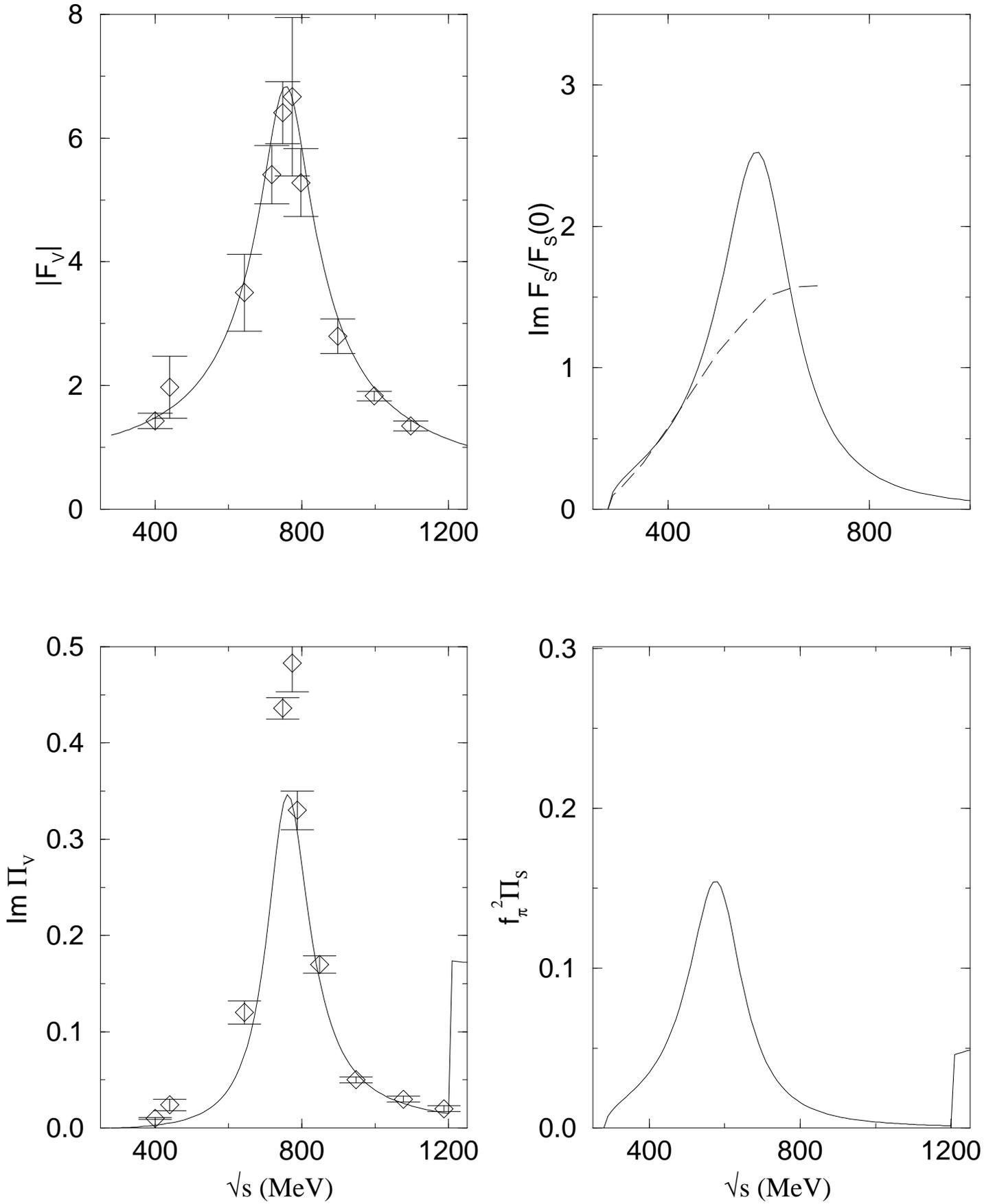
\item Fig.7. The form factors for the one resonance fit of the master
formula approach. Data for ${\bf F}_V$ and ${\bf\Pi}_V$ is from
\cite{exp} (see \cite{calc} for details) and the estimate for
${\bf F}_S$ from a dispersive analysis \cite{au}.
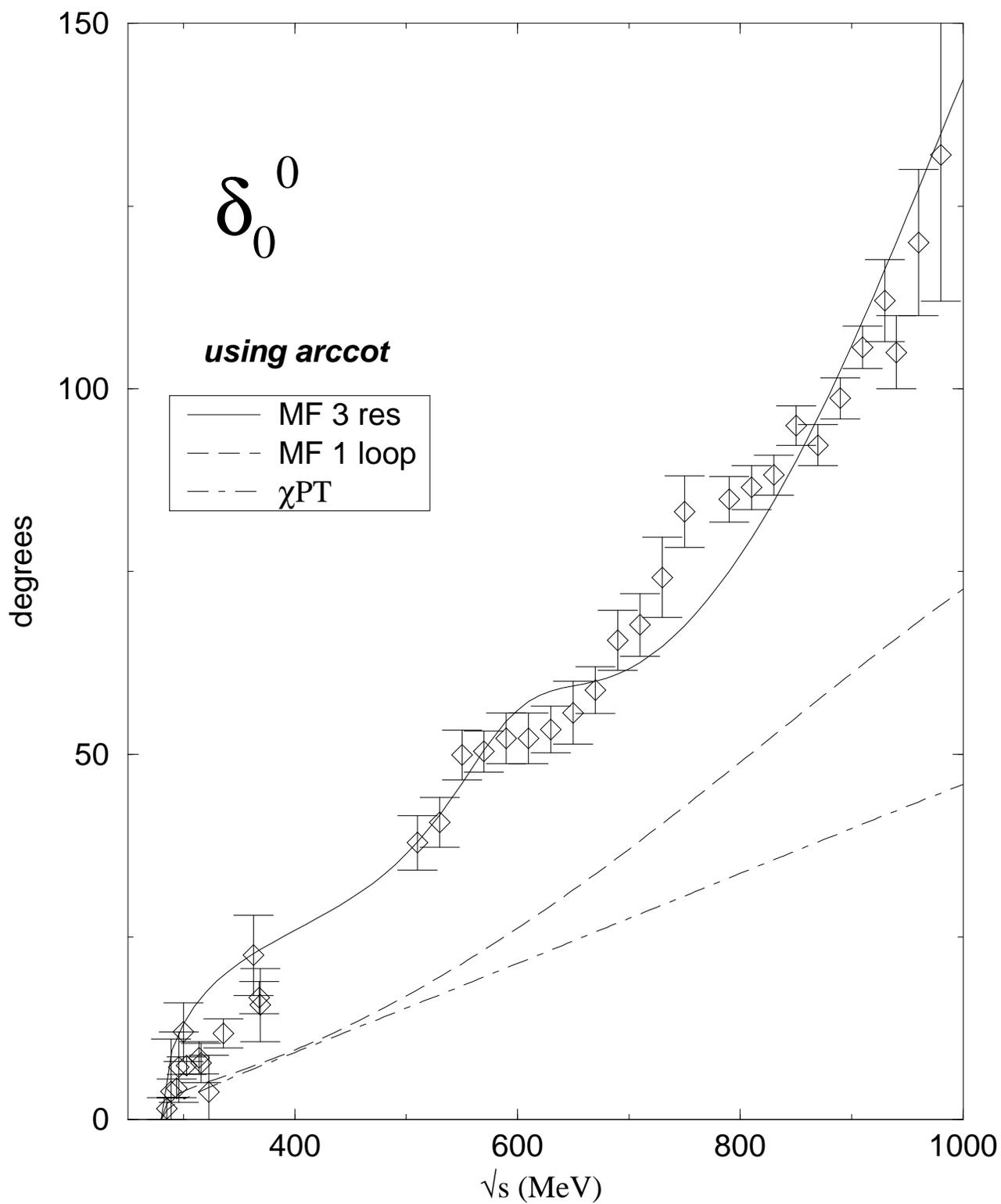
\item Fig.8. Fitting the $\delta_0^0$ to three resonances in the master
equation approach. Also plotted are the master formula to one loop
(dashed line) and $\chi$PT (dotted-dashed line). Data is from
\cite{data,esta}.
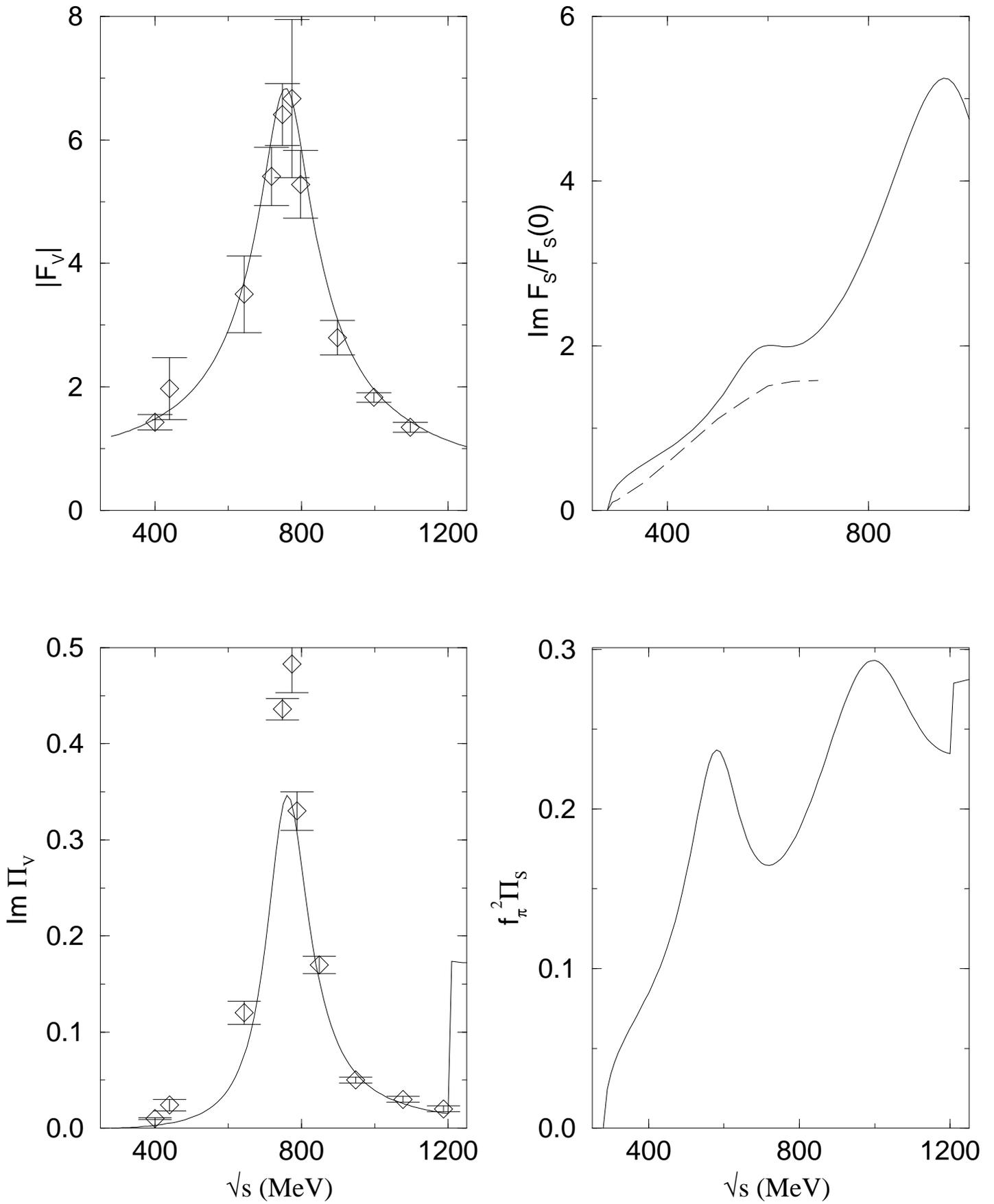
\item Fig.9. The form factors for the three resonance fit to the
master formula (solid line). Other data is as in figure 7.
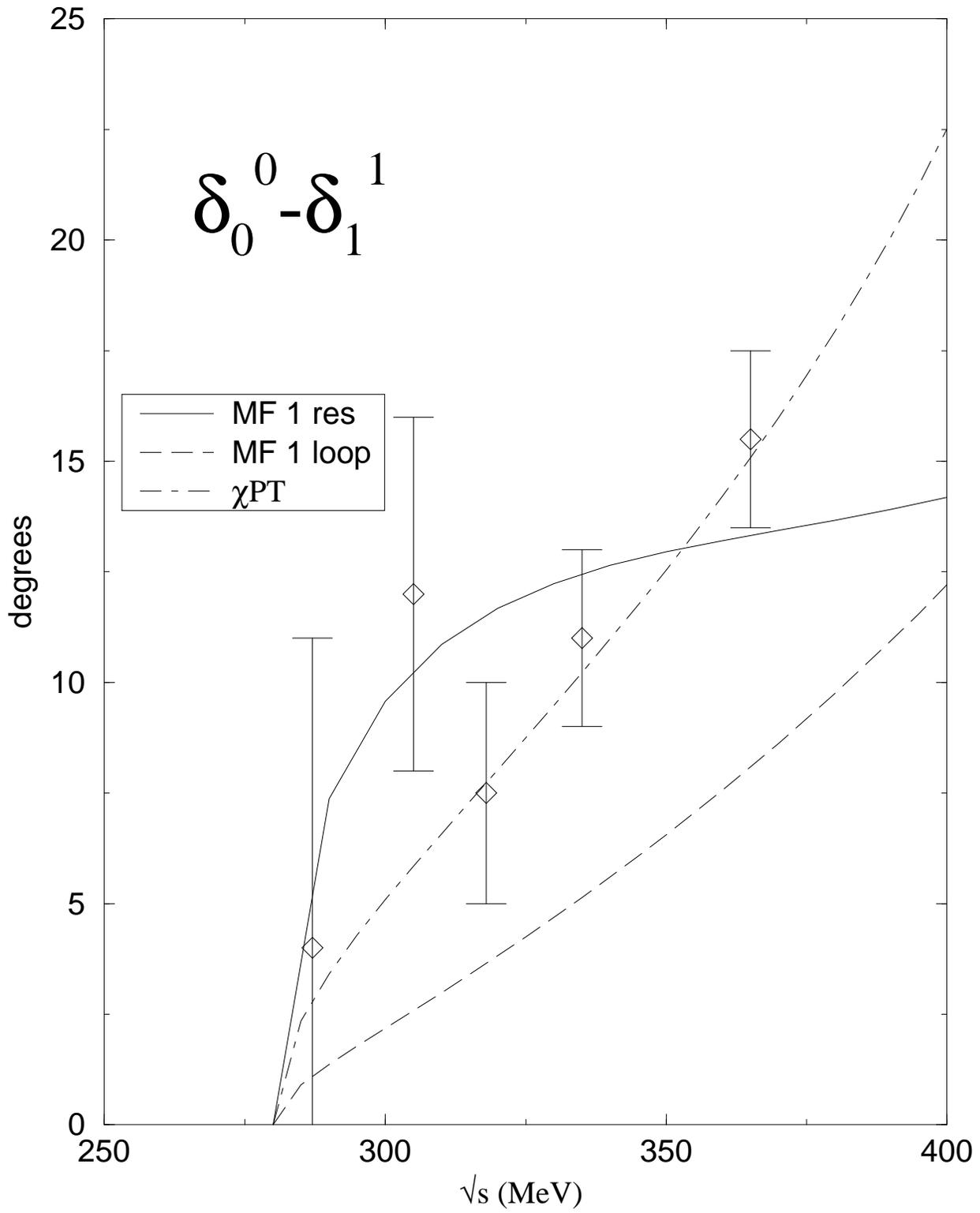
\item Fig.10. $\delta_0^0-\delta_1^1$ with the one resonance fit for the
master formula result (solid line), the one loop master formula
result (dashed line), and $\chi$PT (dotted-dashed line). Data is from
\cite{data}.
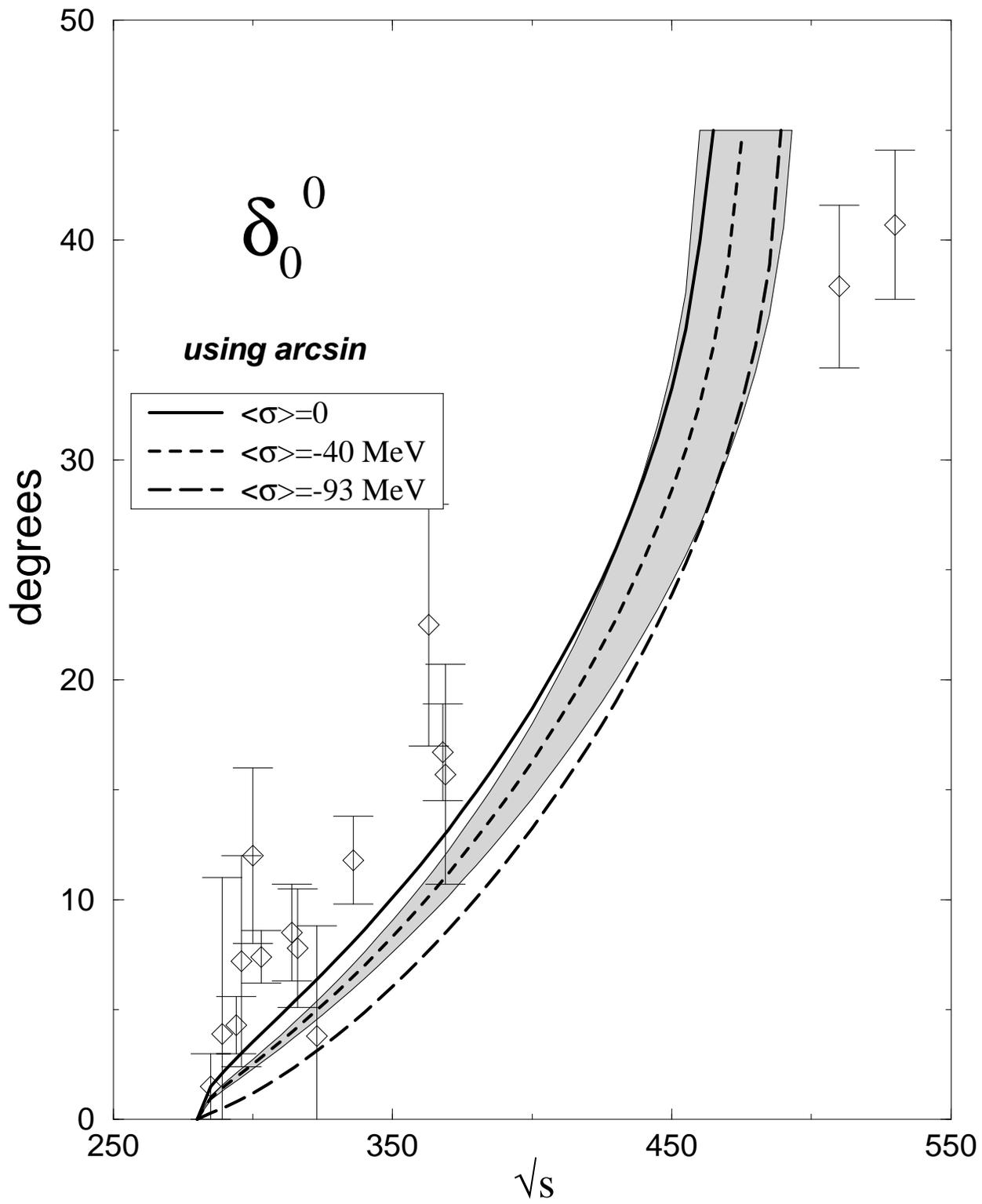
\item Fig.11. The $\delta_0^0$ phase shift for the one loop expansion
of the master formula using $\langle{\hat\sigma}\rangle=0$ (solid
line), $-40$ MeV (dotted line), and $-93$ MeV (dashed line). These
values correspond to $\langle\bar{q}q\rangle=-(290 \,\mbox{MeV})^3$,
$-(240\, \mbox{MeV})^3$, and $0$ respectively. Data is from
\cite{data,esta}.
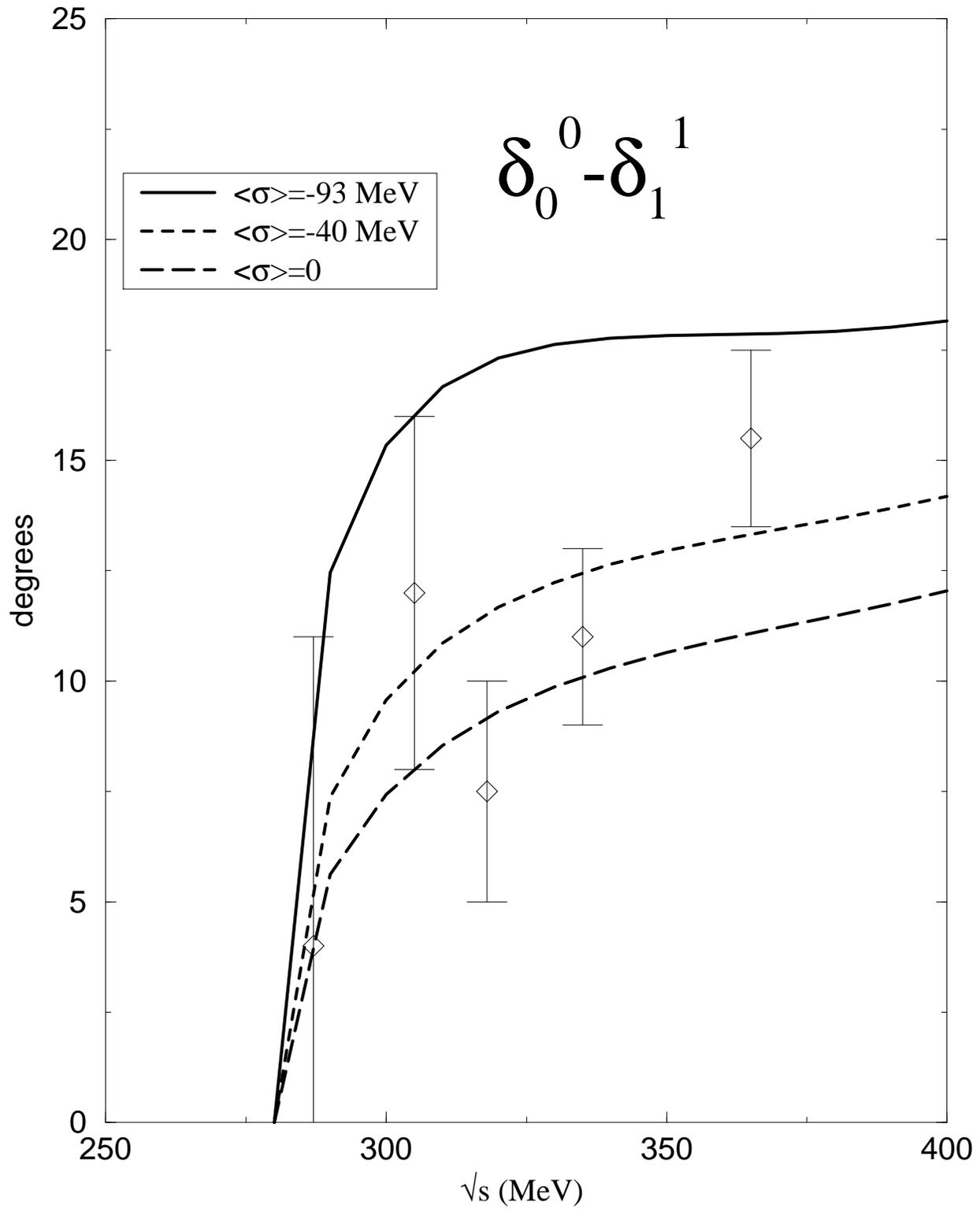
\item Fig.12. $\delta_0^0-\delta_1^1$ for the master formula one
resonance fit using $\langle{\hat\sigma}\rangle=-93$ MeV (solid line),
$-40$ MeV (dashed line), and $0$ (dotted-dashed line). These values
correspond to $\langle\bar{q}q\rangle=0$, $-(240 \,\mbox{MeV})^3$, and
$-(290 \,\mbox{MeV})^3$ respectively. Data is from \cite{data}.
\end{itemize}

\begin{table}[p] Table 1: {Comparison of the experimental
scattering lengths and range parameters to those predicted by the
tree, $\chi$PT, and one loop master formula calculations. For the
master formula, the values $c_1=0.031$ and $16\pi^2{\hat c}_1 =2.46$
were used. Data is from \cite{peter} and $\chi$PT results are from
\cite{gas1,gas2}.}\par
\vskip 2cm
\hspace{3cm}
\begin{tabular}{|r|r|r|r|r|} \hline &&&& \\
 & Experiment & Tree & $\chi$PT & Master \\ \hline\hline &&&& \\
$a_0^0$ & $0.26\pm0.05$ & $0.16$ & $0.20$ & $0.20$ \\ \hline &&&& \\
$m_\pi^2 \,b_0^0$ & $0.25\pm0.03$ & $0.18$  & $0.26$ & $0.24$ \\ \hline
&&&& \\
$m_\pi^2 \,a_1^1$ & $0.038\pm0.002$ & $0.030$ & $0.033$ & $0.038$ \\
\hline &&&& \\
$a_0^2$ & $-0.019\pm0.021$ & $-0.045$ & $-0.041$ & $-0.041$ \\ \hline
&&&& \\
$m_\pi^2 \,b_0^2$ & $-0.082\pm0.008$ & $-0.089$ & $-0.065$ & $-0.068$ \\
\hline
\end{tabular}
\end{table}

\begin{table}[p] Table 2: {The parameters used for the three resonance
fit of the master formal approach. All constants are in MeV except for
$\alpha$, $\beta$, $\gamma$, and the $h_i$'s which are
dimensionless.}\par
\vskip 2cm
\hspace{3cm}
\begin{tabular}{|l|r||l|r||l|r|} \hline &&&&& \\
$m_1$ & 580 & $f_1$ & 125 & $\Gamma_1$ & 175 \\ \hline &&&&& \\
$m_2$ & 980 & $f_2$ & 300 & $\Gamma_2$ & 400 \\
\hline &&&&& \\
$m_3$ & 1300 & $f_3$ & 300 & $\Gamma_3$ & 400 \\ \hline &&&&& \\
${\hat m}$ & 7 & $h_1$ & $1$ & $\langle{\hat\sigma}\rangle$ &
 $-40$ \\ \hline &&&&& \\
$\sqrt{s_0}$ & 1200 & $h_2$ & $13$ & $\alpha$, $\gamma$ &
$-0.17$,\quad$0.3$  \\ \hline
&&&&& \\
$f_\rho$ & 140 & $h_3$ & $-11$ & $\beta$ & $-1.6\times10^{-3}$ \\
\hline
\end{tabular}
\end{table}

\newpage
\eject
\setlength{\baselineskip}{15pt}

\end{document}